\documentclass[aps,prb,a4paper,preprint,showpacs]{revtex4}
\usepackage{graphicx,graphics,color,epsfig}
\usepackage{amsmath}
\usepackage{amsfonts}
\usepackage{amssymb}
\usepackage{slashbox}

\begin{document}
\title{Enhanced spin injection efficiency in a four-terminal double quantum dot system}
\author{Ling Qin,$^{1}$ Hai-Feng L\"{u},$^{2}$ and Yong Guo$^{1,a)}$}
\affiliation{$^{1}$Department of Physics and Key Laboratory of
Atomic and Molecular NanoSciences, Ministry of Education, Tsinghua
University, Beijing 100084, People's Republic of China\\
$^{2}$Department of Applied Physics, University of Electronic
Science and Technology of China, Chengdu 610054, People's Republic
of China}

\begin{abstract}
Within the scheme of quantum rate equations, we investigate the
spin-resolved transport through a double quantum dot system with
four ferromagnetic terminals. It is found that the injection
efficiency of spin-polarized electrons can be significantly
improved compared with single dot case. When the magnetization in
one of four ferromagnetic terminals is antiparallel with the other
three, the polarization rate of the current through one dot can be
greatly enhanced, accompanied by the drastic decrease of the
current polarization rate through the other one. The mechanism is
the exchange interaction between electrons in the two quantum
dots, which can be a promising candidate for the improvement of
the spin injection efficiency.
\end{abstract}

\date{\today}
\maketitle \vskip2pc \narrowtext

\section{introduction}

How to improve the injection efficiency of spin-polarized
electrons from a ferromagnetic (FM) contact into a semiconductor
microstructure has puzzled the researchers in the field of
spintronics for many years.\cite{Zut04} Due to the mismatch of
conductivity between FM metal and semiconductor, spin polarization
is almost lost at the interface,\cite{Sch00} and spin injection
efficiency is very low.\cite{Ham99,Mon98,Fil00,Zhu01} To now,
various ideas have been proposed to solve this problem.
Rashba\cite{Ras00} suggested that tunnel contacts can dramatically
increase spin injection efficiency, which was supported by
subsequent theoretical works.\cite{Fer01,Smi01,Joh03,Tak03} Jiang
et al.\cite{Jia05} demonstrated that the spin injection efficiency
could be improved dramatically by inserting a MgO tunnel barrier
between the ferromagnetic contact and the semiconductor. Optical
injection of spin-polarized carriers across a mismatched
heterostructure is an effective method. By using circular
polarized excitation and detection, it has been demonstrated that
the injected spin-polarized carriers are quite robust and maintain
their polarization memory even after passing through a dense array
of misfit dislocations.\cite{Fie99,Ohn99,Han02,Gha01} However, it
is still desirable to establish electrical, rather than optical,
methods to achieve effective spin injection.

In strongly-correlated electron systems, spin dipole-dipole
interactions between electrons play important roles, which determine
the systems' magnetism, specific heat, and other ground-state
properties. In the weak coupling and strong Coulomb repulsion
regime, the Heisenberg-type exchange interaction
$J\textbf{S}_{1}\cdot \textbf{S}_{2}$ can be derived through
perturbation analysis (e.g., Schrieffer-Wolf transformation). For
electronic transport in mesoscopic systems, electronic spin
correlation drastically affects the conductance and the current
correlation.\cite{Bus00,Don02,Kau06,Chu07,Chu08,Fra07,Tol07,Koe07}
For instance, the double quantum dot (QD) system enables the
realization of the two-impurity Kondo problem, in which a
competition between Kondo correlation and antiferromagnetic
impurity-spin correlation leads to a quantum critical
phenomenon.\cite{Lop02} For the case of spin-polarized transport,
the polarized spin in one dot behaves like an effective magnetic
field and affects the spin transport in another dot through indirect
spin-spin interaction between two dots.\cite{Lu08} Therefore, it is
expected that exchange interaction can induce efficient spin
injection in QD systems.

In this work we propose an electrical and internal scheme to
improve the spin injection efficiency based on a double quantum
dot system, where each dot is connected with two FM electrodes.
Two different configurations are examined, one is the
magnetizations of four FM electrodes are parallel with each other,
and the other is one of them has antiparallel magnetization with
other three ones. We find that in the latter case, due to the
exchange interaction between electrons in the double dot, the
spin-polarization rate of the current through one dot is greatly
enhanced, while the spin-polarization rate through the other one
is drastically suppressed. As for the case of two parallel and two
antiparallel, spin-down electrons can hardly occupy the two dots,
while the spin-up ones dominate in both of the two dots during
transport processes, thus the exchange interaction cannot greatly
enhance the current polarization.

\section{model and formula}

The structure is depicted in Fig. 1. Dot $i$ ($i=$1,2) is
connected to FM leads $i$L and $i$R. The magnetizations of leads
1L, 2L, and 2R are parallel, while that of lead 1R can be parallel
or antiparallel with the other three. We model this system with
the Hamiltonian $H=H_{lead}+H_{dot}+H_{T}$. The FM leads are
described by the Hamiltonian $H_{lead}=\sum\limits_{i\alpha
k\sigma}\varepsilon_{i\alpha k\sigma}a_{i\alpha
k\sigma}^{\dagger}a_{i\alpha k\sigma}$, where $a_{i\alpha
k\sigma}^{\dagger}$ ($a_{i\alpha k\sigma}$) is the creation
(annihilation) operator for electrons with wave vector $k$ in lead
$i\alpha $, $\alpha=$L,R. The isolated double dot are described by
$H_{dot}=\sum\limits_{i\sigma}\varepsilon_{i}d_{i\sigma}^{\dagger}d_{i\sigma}+\sum\limits_{i}U_{i}n_{i\uparrow}n_{i\downarrow}+J\textbf{S}_{1}\cdot
\textbf{S}_{2}$. Here $d^{\dagger}_{i\sigma}$ ($d_{i\sigma}$) is
the creation (annihilation) operator for electrons with spin
$\sigma$ in dot $i$,
$n_{i\sigma}=d^{\dagger}_{i\sigma}d_{i\sigma}$ is the occupation
operator, and $U_{i}$ stands for the intradot Coulomb repulsion.
The last term denotes the Heisenberg exchange coupling with the
exchange coupling parameter $J$ and the spin operator
$\textbf{S}_{i}=(\hbar/2)\sum\limits_{\sigma\sigma'}d^{\dagger}_{i\sigma}$\mbox{\boldmath$\sigma$}$_{\sigma\sigma'}d_{i\sigma'}$.
For simplicity, we neglect the direct interdot tunneling and
interdot Coulomb repulsion.\cite{Lop02,Fra07,Lu08} The tunneling
Hamiltonian between dots and leads is $H_{T}=\sum\limits_{i\alpha
k\sigma}(V_{i\alpha k\sigma}a^{\dagger}_{i\alpha
k\sigma}d_{i\sigma}+\textrm{H.c.})$. In the following, we assume
the coupling coefficient $V_{i\alpha k\sigma}$ to be independent
of $k$ and $U_{1},U_{2}\rightarrow\infty$, thus the double
occupation of each dot is forbidden.

Since the exchange interaction is considered, it is natural to
describe the double dot system by triplet and singlet states,
which are defined as
$|T_{\uparrow}\rangle=|\uparrow\rangle_{1}|\uparrow\rangle_{2}$,
$|T_{\downarrow}\rangle=|\downarrow\rangle_{1}|\downarrow\rangle_{2}$,
$|T_{0}\rangle=(1/\sqrt{2})(|\uparrow\rangle_{1}|\downarrow\rangle_{2}+|\downarrow\rangle_{1}|\uparrow\rangle_{2})$
(triplet states), and
$|S\rangle=(1/\sqrt{2})(|\uparrow\rangle_{1}|\downarrow\rangle_{2}-|\downarrow\rangle_{1}|\uparrow\rangle_{2})$
(singlet state). Following the procedure in previous
works,\cite{Don04,Qin08} we use nine slave-boson operators to
represent these Dirac brackets:
$e^{\dagger}=|0\rangle_{1}|0\rangle_{2}$,
$f^{\dagger}_{1\sigma}=|\sigma\rangle_{1}|0\rangle_{2}$,
$f^{\dagger}_{2\sigma}=|0\rangle_{1}|\sigma\rangle_{2}$,
$d^{\dagger}_{T_{\sigma}}=|T_{\sigma}\rangle$,
$d^{\dagger}_{T_{0}}=|T_{0}\rangle$, and
$d^{\dagger}_{S}=|S\rangle$. Thus,
$d_{i\sigma}=e^{\dagger}f_{i\sigma}+\sigma f^{\dagger}_{\bar{i}\sigma}
d_{T_{\sigma}}+(1/\sqrt{2})\sigma f^{\dagger}_{\bar{i}\bar{\sigma}}[d_{T_{0}}+(-1)^{i}\bar{\sigma}
d_{s}]$ and
$H_{dot}=\sum\limits_{i\sigma}\varepsilon_{i}f_{i\sigma}^{\dagger}
f_{i\sigma}+(\varepsilon_{1}+\varepsilon_{2}+J/4)\sum\limits_{\gamma=\uparrow,\downarrow,0}d^{\dagger}_{T_{\gamma}}d_{T_{\gamma}}+(\varepsilon_{1}+\varepsilon_{2}-3J/4)d^{\dagger}_{S}d_{S}$
with $\bar{1}(\bar{2})=2(1)$ and
$\bar{\uparrow}(\bar{\downarrow})=\downarrow(\uparrow)$.

Using equation of motion, one can derive the dynamical equations
of elements of the density matrix.\cite{Don04} Their statistical
expectations involve the time-diagonal parts of the less Green's
functions, which can be calculated with the help of the Langreth
analytic continuation rules and the Fourier transformation.
Submitting the uncoupled dot's Green's function into the
equations, the mater equations describe the electronic transport
can be derived  as
\begin{eqnarray}
\dot{\hat{\rho}}_{0}&=&\sum_{i\alpha
\sigma}\Gamma^{\sigma}_{i\alpha}\displaystyle\big\{[1-f_{i\alpha}(\varepsilon_{i})]\rho_{i\sigma}-f_{i\alpha}(\varepsilon_{i})\rho_{0}\displaystyle\big\},\nonumber\\
\dot{\hat{\rho}}_{i\sigma}&=&\sum_{\alpha}\displaystyle\bigg\{\Gamma^{\sigma}_{i\alpha}f_{i\alpha}(\varepsilon_{i})\rho_{0}
-\displaystyle\big\{\Gamma^{\sigma}_{i\alpha}[1-f_{i\alpha}(\varepsilon_{i})]+\Gamma^{\sigma}_{\bar{i}\alpha}f_{\bar{i}\alpha}(\varepsilon_{\bar{i}}+J/4)\nonumber\\
&&+\frac{1}{2}\Gamma^{\bar{\sigma}}_{\bar{i}\alpha}[f_{\bar{i}\alpha}(\varepsilon_{\bar{i}}+J/4)+f_{\bar{i}\alpha}(\varepsilon_{\bar{i}}-3J/4)]
\displaystyle\big\}\rho_{i\sigma}+\Gamma^{\sigma}_{\bar{i}\alpha}[1-f_{\bar{i}\alpha}(\varepsilon_{\bar{i}}+J/4)]\rho_{T_{\sigma}}\nonumber\\
&&+\frac{1}{2}\Gamma^{\bar{\sigma}}_{\bar{i}\alpha}[1-f_{\bar{i}\alpha}(\varepsilon_{\bar{i}}+J/4)]\rho_{T_{0}}
+\frac{1}{2}\Gamma^{\bar{\sigma}}_{\bar{i}\alpha}[1-f_{\bar{i}\alpha}(\varepsilon_{\bar{i}}-3J/4)]\rho_{S}\nonumber\\
&&+(-1)^{i}\frac{\bar{\sigma}}{2}\Gamma^{\bar{\sigma}}_{\bar{i}\alpha}[1-\frac{1}{2}f_{\bar{i}\alpha}(\varepsilon_{\bar{i}}+J/4)
-\frac{1}{2}f_{\bar{i}\alpha}(\varepsilon_{\bar{i}}-3J/4)](\rho_{S,T_{0}}+\rho_{T_{0},S})\displaystyle\bigg\},\nonumber\\
\dot{\hat{\rho}}_{T_{\sigma}}&=&\sum_{i\alpha}\Gamma^{\sigma}_{i\alpha
}\displaystyle\big\{f_{i\alpha}(\varepsilon_{i}+J/4)\rho_{\bar{i}\sigma}-[1-f_{i\alpha}(\varepsilon_{i}+J/4)]\rho_{T_{\sigma}}\displaystyle\big\},\nonumber\\
\dot{\hat{\rho}}_{T_{0}}&=&\frac{1}{2}\sum_{i\alpha\sigma}\Gamma^{\sigma}_{i\alpha
}\displaystyle\big\{f_{i\alpha}(\varepsilon_{i}+J/4)\rho_{\bar{i}\bar{\sigma}}-[1-f_{i\alpha}(\varepsilon_{i}+J/4)]\rho_{T_{0}}\nonumber\\
&&+\frac{1}{4}(-1)^{i}\sigma[1-\frac{1}{2}f_{i\alpha}(\varepsilon_{i}+J/4)-\frac{1}{2}f_{i\alpha}(\varepsilon_{i}-3J/4)](\rho_{S,T_{0}}+\rho_{T_{0},S})\displaystyle\big\},\nonumber\\
\dot{\hat{\rho}}_{S}&=&\frac{1}{2}\sum_{i\alpha\sigma}\Gamma^{\sigma}_{i\alpha
}\displaystyle\big\{f_{i\alpha}(\varepsilon_{i}-3J/4)\rho_{\bar{i}\bar{\sigma}}-[1-f_{i\alpha}(\varepsilon_{i}-3J/4)]\rho_{S}\nonumber\\
&&+\frac{1}{4}(-1)^{i}\sigma[1-\frac{1}{2}f_{i\alpha}(\varepsilon_{i}+J/4)-\frac{1}{2}f_{i\alpha}(\varepsilon_{i}-3J/4)](\rho_{S,T_{0}}+\rho_{T_{0},S})\displaystyle\big\},\nonumber\\
\dot{\hat{\rho}}_{T_{0},S}&=&\frac{1}{4}\sum_{i\alpha\sigma}(-1)^{i}\sigma\Gamma^{\sigma}_{i\alpha}
\displaystyle\big\{[1-f_{i\alpha}(\varepsilon_{i}+J/4)]\rho_{T_{0}}+[1-f_{i\alpha}(\varepsilon_{i}-3J/4)]\rho_{S}\nonumber\\
&&-[f_{i\alpha }(\varepsilon_{i}+J/4)+f_{i\alpha}(\varepsilon_{i}-3J/4)]\rho_{\bar{i}\bar{\sigma}}\displaystyle\big\}\nonumber\\
&&+\displaystyle\big\{iJ-\frac{1}{2}\sum_{i\alpha\sigma}\Gamma^{\sigma}_{i\alpha
}[1-\frac{1}{2}f_{i\alpha}(\varepsilon_{i}+J/4)-\frac{1}{2}f_{i\alpha}(\varepsilon_{i}-3J/4)]\displaystyle\big\}\rho_{T_{0},S},
\end{eqnarray}
where the elements of the density matrix are defined as
$\hat{\rho}_{0}=e^{\dagger}e$,
$\hat{\rho}_{i\sigma}=f^{\dagger}_{i\sigma}f_{i\sigma}$,
$\hat{\rho}_{T_{\gamma}}=d^{\dagger}_{T_{\gamma}}d_{T_{\gamma}}$,
and $\hat{\rho}_{S}=d^{\dagger}_{S}d_{S}$. These elements represent
the probability that both dots are empty, one electron with spin
$\sigma$ occupies dot $i$, and two electrons form the triplet states
and the singlet state, respectively. They satisfy the completeness
relation
$\rho_{0}+\sum\limits_{\sigma}(\rho_{1\sigma}+\rho_{2\sigma}+\rho_{T_{\sigma}})+\rho_{T_{0}}+\rho_{S}=1$.
$\rho_{S,T_{0}}$ is induced by the exchange interaction. $f_{i\alpha
}(\omega)=[1+e^{(\omega-\mu_{i\alpha})/k_{B}T}]^{-1}$ is the Fermi
distribution function of lead $i\alpha$, and
$\Gamma^{\sigma}_{i\alpha }=\sum\limits_{k}2\pi|V_{i\alpha
k\sigma}|^{2}\delta(\omega-\varepsilon_{i\alpha k\sigma})$ is the
coupling strength between lead $i\alpha $ and dot $i$. In the
stationary situation, the elements of the density matrix can be
derived, and the spin component of current in lead $i\alpha$ can
be obtained as
\begin{eqnarray}
I^{\sigma}_{i\alpha
}&=&\frac{e}{\hbar}\Gamma^{\sigma}_{i\alpha}\displaystyle\big\{f_{i\alpha}(\varepsilon_{i})\rho_{0}-[1-f_{i
\alpha}(\varepsilon_{i})]\rho_{i\sigma}+f_{i\alpha}(\varepsilon_{i}+J/4)\rho_{\bar{i}\sigma}+\frac{1}{2}[f_{i
\alpha}(\varepsilon_{i}+J/4)\nonumber\\
&&+f_{i\alpha}(\varepsilon_{i}-3J/4)]\rho_{\bar{i}\bar{\sigma}}
-[1-f_{i\alpha}(\varepsilon_{i}+J/4)]\rho_{T_{\sigma}}-\frac{1}{2}[1-f_{i
\alpha}(\varepsilon_{i}+J/4)]\rho_{T_{0}}\nonumber\\
&&-\frac{1}{2}[1-f_{i\alpha}(\varepsilon_{i}-3J/4)]\rho_{S}
+(-1)^{i}\frac{\sigma}{2}[1-\frac{1}{2}f_{i\alpha}(\varepsilon_{i}+J/4)\nonumber\\
&&-\frac{1}{2}f_{i\alpha}(\varepsilon_{i}-3J/4)](\rho_{S,T_{0}}+\rho_{T_{0},S})\displaystyle\big\}.
\end{eqnarray}
 When
$J\rightarrow0$, these quantum rate equations reduce to the
equations describing two separate dots.\cite{Sou08,Bul99} For a single dot,
interplay between Coulomb interaction and spin accumulation in
the dot can result in a bias-dependent current polarization, which can be suppressed in the P
alignment and enhanced in the AP case.\cite{Sou08}
Furthermore, the spin flip process make the occupations of spin-up
and spin-down electrons in the dots tend to be equal,
which can weaken the enhancement of current spin-polarization rate.

\section{numerical results and discussions}

For numerical calculations, we choose meV to be the energy unit
and set $k_{B}T=0.002$. The polarization rates of all leads are
assumed to be $P=0.4$, and the coupling strength is
$\Gamma^{\sigma}_{i\alpha}=(1+\sigma P)\Gamma$, except for lead
$1$R it becomes $(1\pm\sigma P)\Gamma$, where $+$ for the parallel
(P) configuration and $-$ for the antiparallel (AP) one. $\Gamma$
and $J$ are set to be 0.01 and 0.2,
respectively,\cite{Hat08,Tol07,Lu08} and the current are
normalized to $e\Gamma/h$. The exchange coupling $J$ between two
dots is the key interaction to improve the spin injection
efficiency. Its strength sensitively depend on the e-e Coulomb
interaction, interdot coupling, Bychkov-Rashba spin-orbit
interaction, and magnetic field. $J$ can reach several hundreds eV
and can be tuned to ferromagnetic ($J < 0$) type in the presence
of magnetic field.\cite{Fqu09} Typical value of the dot-lead
coupling strength $\Gamma$ is order of 1$\mu$eV, therefore,
$J/\Gamma\gg1$, which makes sure that the quantum rate equations
are valid in every bias region.

For clarity, first we show relevant results for single QD system
connected to two FM leads.\cite{Sou08} The spin components of the
current are
$I^{\sigma}=(e/h)(\Gamma^{\sigma}_{L}\Gamma^{\uparrow}_{R}
\Gamma^{\downarrow}_{R})/(\Gamma^{\uparrow}_{L}
\Gamma^{\downarrow}_{R}+\Gamma^{\downarrow}_{L}
\Gamma^{\uparrow}_{R}+\Gamma^{\uparrow}_{R}\Gamma^{\downarrow}_{R})$.
Thus, the spin-polarization rate is
$\eta=(I^{\uparrow}-I^{\downarrow})/(I^{\uparrow}+I^{\downarrow})=P_{L}=P$,
regardless of whether the system is in P or AP
configuration.\cite{Sou08,Sou07} However, for the four-terminal
structure, when the exchange interaction is absent,
$n_{1\sigma}=n_{2\sigma}=1/3$ for the P configuration, while
$n_{1\uparrow}>n_{1\downarrow}$ and
$n_{2\uparrow}=n_{2\downarrow}$ for the AP one. Since the exchange
interaction is sensitive to the spin-dependent occupation numbers
in the two dots, we expect that in the P configuration the
exchange interaction has little influence on the current
polarization, while in the AP one it can affect the transport
properties greatly. Further, we apply a large bias between leads
1L and 1R to make sure that $\varepsilon_{1}$ is deeply in the
bias window.

Fig. 2(a) shows variations of $I^{\sigma}_{2}$ and $n_{2\sigma}$
with the bias voltage in the P configuration. In the following,
$I^{\sigma}_{2}$ is denoted by $I^{\sigma}$, for convenience.  As
expected, both $I^{\uparrow}$ and $I^{\downarrow}$ increase
monotonously with the bias, and three steps occur when $\mu_{2L}$
crosses $\varepsilon_{2}-3J/4$, $\varepsilon_{2}$, and
$\varepsilon_{2}+J/4$, respectively. They correspond to the
situations that electrons tunnel through dot 2 via the singlet
state, the energy level $\varepsilon_{2}$, and the triplet states.
Here we mark the bias regions
$\varepsilon_{2}-3J/4<V/2<\varepsilon_{2}$,
$\varepsilon_{2}<V/2<\varepsilon_{2}+J/4$, and
$V/2>\varepsilon_{2}+J/4$ as I, II, and III, respectively. In each
region, $I^{\uparrow}>I^{\downarrow}$. However, in region I,
$n_{2\downarrow}>n_{2\uparrow}$, which is different from the case
of isolated single dot, where $n_{\uparrow}=n_{\downarrow}$ and
$\eta=P=0.4$. Since $n_{2\downarrow}>n_{2\uparrow}$, $\eta_{2}$ is
suppressed from 0.4, accompanied by the increase of $\eta_{1}$.
When the bias rises beyond region I, both $\eta_{1}$ and
$\eta_{2}$ return to 0.4. So in the P configuration we can not
enhance $\eta_{2}$ from its original value in single dot case.

In the AP configuration, $\eta_{2}$ can be strongly modified from
the single dot case by the exchange interaction (see Fig. 3).
Figs. 3(a) indicates both $I^{\uparrow}$ and $I^{\downarrow}$
increase monotonously with the bias, which is similar to that in
the P configuration. However, from region I to region III, the
discrepancy between $I^{\uparrow}$ and $I^{\downarrow}$ keeps
increasing, resulting in the enhancement of $\eta_{2}$ in Fig.
3(b). In region III, $\eta_{2}$ approaches 0.7, which is much
larger than its original value 0.4 in single dot system. At the
same time, $\eta_{1}$ keeps decreasing when bias increases from
region I to region III, and finally becomes smaller than 0.1. It
is concluded that in the AP configuration one can greatly enhance
the current polarization rate through one dot, accompanied by
decrease of the current polarization rate through another dot.
Such phenomenon looks as if the current polarization rate is
``transferred" from one circuit to the other.

The enhancement of the current polarization rate can be understood
with the aid of the expression of the current. Due to the absence
of intradot spin flips, both the amplitude and spin polarization
of the total current through dot $2$ are conserved, i.e.,
$I_{2L}^{\sigma}= I_{2R}^{\sigma}$. For simplicity, the current
$I_{2R}^{\sigma}$ is chosen in the calculation because it has an
uniform expression in all three regions:
$I^{\sigma}=(e/h)\Gamma^{\sigma}_{2R}[\rho_{2\sigma}+\rho_{T_{\sigma}}+(1/2)\rho_{T_{0}}+(1/2)\rho_{S}]$.
The first term denotes the process that one electron tunnels
through dot 2 via the energy level $\varepsilon_{2}$, and the
second to fourth terms denote the processes that one electron with
spin $\sigma$ transports through dot 2 via the triplet states and
the singlet state. Because in $T_{0}$ and $S$ states, electrons
with spin $\sigma$ or $\bar{\sigma}$ have the same probability to
occupy dot 2, both the third and the fourth terms have a factor
$1/2$. From Fig. 3(b), in region I we can see $\eta_{2}$ is
slightly larger than $P=0.4$. In this region, only the energy
level $\varepsilon_{2}-3J/4$ enters the bias window, and electrons
can only form the singlet state, which makes $\rho_{S}$ much
larger than other elements [see Figs. 3(c) and 3(d)]. Thus, the
forth term dominates in expression of the current, and we have
$I^{\sigma}=(e/2h)\Gamma^{\sigma}_{2R}(\rho_{S}+\rho_{T_{0}})$,
$\eta_{2}=({I^{\uparrow}-I^{\downarrow}})/({I^{\uparrow}+I^{\downarrow}})=P=0.4$.
When the effects of $\rho_{2\sigma}$ and $\rho_{T_{\sigma}}$ are
considered, the value of $\eta_{2}$ is slightly modified. From Eq.
(1) we can obtain $\rho_{2\sigma}
\approx\Gamma^{\bar{\sigma}}_{1R}\rho_{S}/[2(\Gamma^{\sigma}_{1L}+\Gamma^{\bar{\sigma}}_{1L}+\Gamma^{\sigma}_{2L}+\Gamma^{\sigma}_{2R})]$.
Here we denote $(1+\sigma P)\Gamma=\Gamma^{\sigma}$, then
$\Gamma^{\sigma}_{i\alpha}=\Gamma^{\sigma}$, except for
$\Gamma^{\sigma}_{1R}=\Gamma^{\bar{\sigma}}$. Thus,
$\rho_{2\uparrow}\approx\rho_{S}/[2(3+\Gamma^{\downarrow}/\Gamma^{\uparrow})]>\rho_{2\downarrow}\approx\rho_{S}/[2(3+\Gamma^{\uparrow}/\Gamma^{\downarrow})]$,
and $\eta_{2}$ is enhanced from 0.4, as shown in Fig. 3(b). In
region I, $\rho_{S}$ is much larger than other elements, which
means that during most of the time electrons in the double dot
form the singlet state. So $\rho_{2\sigma}$ is mainly contributed
by the process that an electron in dot 1 tunnels to lead 1R and
breaks the singlet state. Noticing that in such a configuration,
$\Gamma^{\downarrow}_{1R}>\Gamma^{\uparrow}_{1R}$, electron with
spin $\downarrow$ can tunnel to lead 1R more easily, and left an
electron with spin $\uparrow$ in dot 2, which makes
$\rho_{2\uparrow}>\rho_{2\downarrow}$.

When the bias locates in region II, the direct tunneling channel
at $\varepsilon_{2}$ opens. We can see the enhancement of
$\rho_{2\uparrow}$ ($\rho_{T_{\uparrow}}$) is larger than
$\rho_{2\downarrow}$ ($\rho_{T_{\downarrow}}$), which results in
further increase of $\eta_{2}$. Here
$\rho_{2\sigma}=[\Gamma^{\sigma}_{2L}\rho_{0}+\Gamma^{\sigma}_{1R}\rho_{T_{\sigma}}
+(1/2)\Gamma^{\bar{\sigma}}_{1R}(\rho_{T_{0}}+\rho_{S})]/(\Gamma^{\sigma}_{1L}+\Gamma^{\bar{\sigma}}_{1L}+\Gamma^{\sigma}_{2R})$.
It is obvious that the increase of $\rho_{2\sigma}$ is mainly
owing to the term $\Gamma^{\sigma}_{2L}\rho_{0}$ in the numerator,
which is absent in region I. Following the same procedure, this
term reads
$\Gamma^{\sigma}_{2L}\rho_{0}/(\Gamma^{\sigma}_{1L}+\Gamma^{\bar{\sigma}}_{1L}+\Gamma^{\sigma}_{2R})=\rho_{0}/(2+\Gamma^{\bar{\sigma}}/\Gamma^{\sigma})$,
so the increase of $\rho_{2\uparrow}$ is larger than that of
$\rho_{2\downarrow}$, and $\eta_{2}$ is enhanced from its value in
region I.

When the bias enters region III, $\rho_{2\sigma}$ and
$\rho_{T_{\sigma}}$ keep increasing, and the enhancement of
$\rho_{T_{\uparrow}}$ is much more than other elements. This is
because now the channel at $\varepsilon_{2}+J/4$ opens, and if dot
1 is occupied, electrons in lead 2L can directly tunnel into dot 2
and form the triplet state $T_{\sigma}$. Since lead 1R is in
antiparallel with lead 1L, in most of the time, dot 1 is occupied
by one electron with spin $\uparrow$. As a consequence, electrons
with spin $\uparrow$ in lead 2L is more available to tunnel into
dot 2 and form the triplet state $T_{\uparrow}$, which makes
$\rho_{T_{\uparrow}}\gg\rho_{T_{\downarrow}}$. This can also be
seen in the formula
$\rho_{T_{\sigma}}=(\Gamma^{\sigma}_{2L}\rho_{1\sigma}+\Gamma^{\sigma}_{1L}\rho_{2\sigma})/(\Gamma^{\sigma}_{1R}+\Gamma^{\sigma}_{2R})$,
where the first term in the numerator makes $\rho_{T_{\uparrow}}$
increase intensively in region III. Thus, $\eta_{2}$ is greatly
enhanced in region III.

In the case of $J/\Gamma\gg1$, the analytical expressions in
region I, II, and III are $\eta_{2}\sim 191P/(165-34P^{2})$,
$120P/(84+5P^{2})$, and $51P/(27+6P^{2})$, respectively. For
$P=0.4$, $\eta_{2} \sim$ 0.454, 0.542, and 0.673, which is
consistent with our numerical results. As expected, when
$P\rightarrow1$, $\eta_{2}\rightarrow1$ in all regions. If we tune
the bias into region III, the injection efficiency can be enhanced
to almost twice of its original value. In the inset of Fig. 3(a),
we present the variations of $\eta_{2}$ with $P$ for different
situations. It can be seen that when $P$ is small, $\eta_{2}$ is
greatly enhanced by the exchange interaction.

\section{conclusions}

In summary, we propose a scheme based on a four-terminal double
quantum dot system to improve the spin injection efficiency
greatly. We find that in the antiparallel configuration, the
spin-polarization rate through one quantum dot can be dramatically
enhanced, while the polarization rate through the other one is
suppressed. The operating mechanism is the exchange interaction
between the two quantum dots.

This project was supported by the NSFC (No. 10774083 and
No.10974109) and by the 973 Program (No. 2006CB605105).

\newpage

\begin{figure}[ht]
\caption{(color online) The system with two quantum dots coupled
to four external FM leads. The magnetizations of three leads are
parallel with each other, while the magnetization of lead 1R can
be parallel (P) or antiparallel (AP) with the other
three.}\label{dispers}
\end{figure}

\begin{figure}[ht]
\caption{(color online) The spin component of the current in dot 2
(a) and the spin-polarization rate (b) versus bias in the P
configuration. The inset in (a) shows the variations of the
occupation numbers in dot 2.}\label{dispers}
\end{figure}

\begin{figure}[ht]
\caption{(color online) The transport properties in the AP
configuration. (a) The spin component of the current versus bias.
Ihe inset shows the variations of the spin-polarization rate with
$P$ in different situations. The solid line corresponds to the
single dot case, and the dashed, dotted, and dash-dotted lines
correspond to the situations that the bias locates in region I,
II, and III, respectively. (b) The spin-polarization versus bias.
(c) and (d) The corresponding elements of the density matrix
versus the bias.}\label{dispers}
\end{figure}

\end{document}